\documentclass[aps,prb, twocolumn]{revtex4-1}
\usepackage{graphicx}% Include figure files
\usepackage{dcolumn}% Align table columns on decimal point
\usepackage{bm}% bold math
\begin{document}

%\preprint{}

\title{Plasmon-phonon coupled modes in graphene tuned by the carrier concentration of the semiconductor substrates}



\author{Lei Wang}
\author{Wei Cai}\email{Corresponding author: weicai@nankai.edu.cn}
%\author{Weiwei Luo}
%\author{Zenghong Ma}
%\author{Chenglin Du}
\author{Xinzheng Zhang}
\author{Jingjun Xu}\email{Corresponding author: jjxu@nankai.edu.cn}
\affiliation{The Key Laboratory of Weak-Light Nonlinear Photonics, Ministry of Education, School of Physics and TEDA Applied Physics Institute, Nankai University, Tianjin 300457, China}


\date{\today}

\begin{abstract}
The interaction between graphene plasmons and surface phonons of a semiconductor substrate is investigated,  which can be efficiently controlled by the carrier injection of the substrate. The energy and lifetime of surface phonons in a substrate depend a lot on the carrier concentration,  which provides a new machanism to tune plasmon-phonon coupled modes (PPCMs). More specifically, the dispersion and lifetime of PPCMs can be controlled by the carrier concentration change of the substrate. The energy of PPCMs for a given momentum increases as the carrier concentration of a substrate increases. On the other hand, the momentum of PPCMs for a given energy decreases when the carrier concentration of the substrate increases. Lifetime of PPCMs is always larger than the intrinsic lifetime of graphene plasmons without plasmon-phonon coupling.

\end{abstract}

\pacs{}
%\keywords{}

\maketitle

\section{Introduction}
Recently, graphene plasmons (GPs) have been focused much attention for their extremely large field enhancement and relatively long propagation length \cite{JBS2009, KCG2011}. For these unique optical properties, graphene is believed as one of the best plasmon materials in the infrared and terahertz regimes.  Due to the large momentum mismatch between free light and graphene plasmons, graphene plasmons cannot be excited by free light along translation invariant directions. Therefore, structured graphene such as graphene ribbons \cite{NGG20112}, periodic ribbon arrays \cite{JGH2011}, graphene disks \cite{FTS2013} and graphene rings \cite{paper9} are put forward to break up translation invariance. Meanwhile, some research groups have used a scattering-type scanning near-field optical microscope (s-SNOM) to generate and mapping graphene plasmons \cite{FAB2011, FRA2012, CBA2012}. However, the generation of highly wavelength controllable and long-lived GPs with a relatively simple experimental method is still the bottleneck for applications of GPs.

It is a known fact that GPs can be actively controlled by bias voltages or chemical doping \cite{GPN2012}. But the former method is usually relatively difficult and costed, and the latter method may lead to additional defect scattering in graphene. Thus it is difficult to obtain high quality graphene. For long wavelengths $k \ll k_{\text{F}}$, the dispersion of graphene plasmons reduces to \cite{KCG2011}
\begin{equation}
 q=\frac{\hbar\epsilon_{\text{eff}}\omega_{pl}^2}{2\alpha cE_{\text{F}}} ,
\label{eqn1}
\end{equation}
where $\omega_{pl}$ is the energy of GPs without plasmon-phonon coupling, $\epsilon_{\text{eff}}=(1+\epsilon_{\infty})/2$ is the effective average over the dielectric constants of the substrate and air, $\alpha \approx$1/137 is the fine structure constant,  and $E_{\text{F}}$ is the Fermi energy linked to the carrier density of graphene. This equation means that the properties of substrates can be an equally important factor as the Fermi energy of graphene to affect the GPs, especially when the coupling between GPs and surface phonons of the substrate is existing \cite{FG2008,HD2010,LW2010,HD2013}. These plasmon-phonon coupled modes (PPCMs) can reduce intrinsic damping of GPs and process longer lifetime than GPs, which have been verified experimentally \cite{LW2010, YLZ2013}. However, the energy of PPCMs usually locates near the surface phonon energy of substrates, which is determined by the phonon of substrates. As a result, these PPCMs can hardly be controlled. In this paper, we propose using a semiconductor material as the substrate of graphene. Through adjusting the coupling between free electron oscillations and phonons of the substrate, one can change the energy and lifetime of the surface phonon, thus realize the actively control of the properties of PPCMs. And the carrier concentration of semiconductor substrates can be easily modified by application of electrical fields or light, therefor these PPCMs can also be controlled by applying external visible or near-infrared optical excitations or electrical fields, which is much simpler than changing the intrinsic property of graphene.

\section{Models and calculation methods}

\begin{figure}[htb]
\includegraphics [width=7cm]{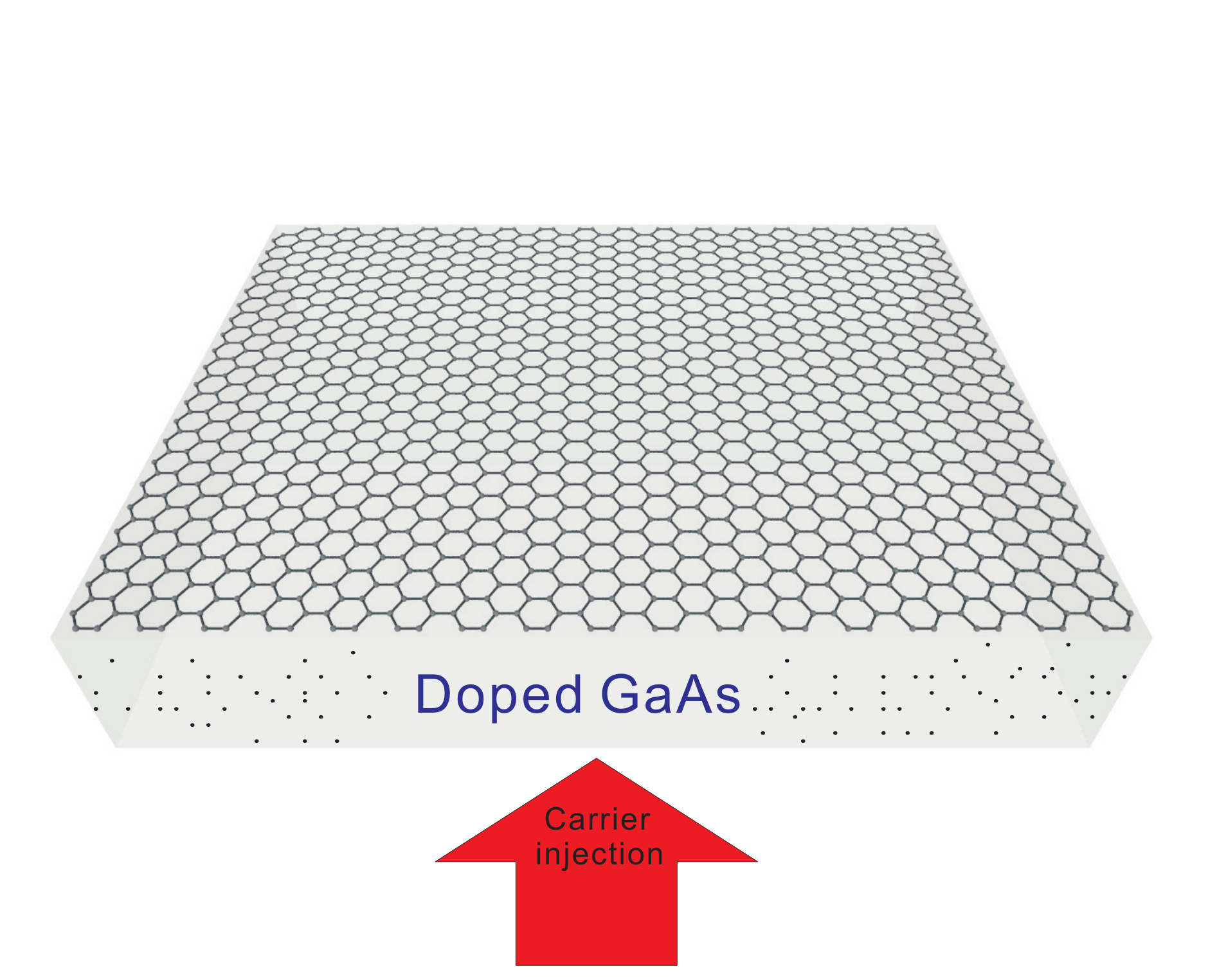}
\caption{The sketch of graphene lay on doped GaAs. The carrier concentration of GaAs can be tuned by field effect or photon excitation. The dots indicate the free electrons injected into GaAs.}
\label{fig1}
\end{figure}

Our proposed scheme is indicated in Fig. \ref{fig1}. A graphene sheet is placed on a doped semiconductor substrate. Gallium arsenide (GaAs), which is a typical semiconductor material, is taken for an example. The carrier concentration of GaAs can be tuned by external electrical fields or optical excitations. The phonons and free electrons included dielectric function of polar semiconductor materials is described as follows \cite{V1965}:
\begin{equation}
 \epsilon (\omega)=\epsilon_{\infty}+\frac{(\epsilon_{0}-\epsilon_{\infty})\omega_{\text{TO}}^2}{\omega_{\text{TO}}^2-\omega^2-i\omega\Gamma}-
 \frac{\epsilon_{\infty}\omega_{p}^2}{\omega^2+i\omega\delta_p},
\label{eqn2}
\end{equation}
where $\omega_{\text{TO}}$ denotes the frequency of transverse optical phonons, $\Gamma$ is the damping rate related to phonons, $\omega_p^2=4\pi ne^2/m^*\epsilon_{\infty}$ denotes the plasma energy of free electrons, $\delta_p=\hbar/\tau_p$ is the damping rate related to the electron scattering in GaAs. $\epsilon_{0}$ and $\epsilon_{\infty}$ are the static and high frequency dielectric constants, respectively. For GaAs, the parameters \cite{LYR2005,B1982} are $\omega_{\text{TO}}=268.7 $ cm$^{-1}$, $\epsilon_{0}=13$, $\epsilon_{\infty}=11$ , $m^*=0.067 m_e$, $\mu_{\text{GaAs}}=8500$ $\text{cm}^2$/Vs. So the reduced parameters $\omega_{\text{LO}}=\sqrt{\epsilon_{\infty}/\epsilon_{0}} \omega_{\text{TO}}=291.1$ $\text{cm}^{-1}$, $\tau_p=\frac{\mu_{\text{GaAs}}m^*}{e}=0.324$ ps. The optical response of graphene is described by the in-plane complex conductivity which is computed with local random phase approximation (RPA) \cite{WSS2006, HD2007}. The Fermi energy of graphene is set as $0.2$ eV (corresponding to carrier density $2.94\times10^{12}$ $\text{cm}^{-2}$), which is a typical value for graphene in the air. With an intrinsic electron relaxation time $\tau_e=\mu E_{\text{F}}/ev_{\text{F}}^2$, where $v_{\text{F}}\approx c/300$ is the Fermi velocity, $\mu=10000$ $\text{cm}^2$/Vs is the measured DC mobility \cite{NGM2004}, and the temperature is set as 300K.

The dispersion relation of graphene between two dielectric materials can be obtained from the pole of $p$-polarized Fresnel reflection coefficients \cite{KCG2011}:
\begin{equation}
\epsilon/\sqrt{\epsilon k_0^2-q^2}+1/\sqrt{k_0^2-q^2}=-4\pi\sigma(\omega)/\omega
\label{eqn3}
\end{equation}
In order to get the lifetime of PPCMs, the coupling mechanism between gaphene and GaAs can be considered as follows. The coupling strength between graphene plasmons and surface polar phonons is given by $M(q)=\hbar g e^{-q z_0} \sqrt{2\pi c\alpha \omega_{so}/q}$, where $g=\sqrt{1/(1+\epsilon_{\infty})-1/(1+\epsilon_{0})}$ is the coupling constant, $\omega_{so}=\sqrt{(\epsilon_0+1)/(\epsilon_{\infty}+1)}\omega_{\text{TO}}$ is the original surface phonon frequency, and $z_0$ is the distance between graphene and substrate, which is set as zero throughout this paper. The exchange potential due to the phonon coupling can be written as  $v_{so}=|M(k)|^2 G^{(0)}(\omega,\tau_{so})$, where $G^{(0)}(\omega,\tau_{so})=\frac{2\omega_p}{\hbar[(\omega+i/\tau_{so})^2-\omega_{so}^2]}$ is the surface phonon propagator\cite{WM1972}. The total effective carrier interaction results from the Coulomb interaction $v_c=2\pi e^2/q\epsilon_{\text{eff}}$, the phonon exchange potential $v_{so}$ and electron-electron interaction potential from free carrier of GaAs $v_p$, it reads\cite{Mahan}:

\begin{equation}
\epsilon_{\text{rpa}}=\epsilon_{\text{eff}}-v_c\Pi^0_g{(q,\omega)}-\epsilon_{\text{eff}}v_{\text{sc-so}}\Pi^0_g{(q,\omega)}-v_c'\Pi^0_s{(q,\omega)}
\label{eqn4}
\end{equation}
where $\Pi^0_g(q,\omega)\approx\frac{E_Fq^2}{\pi\hbar^2\omega(\omega+i\delta_e)}$ is the polarizability of graphene, depending on $\delta_e=\hbar/\tau_e$. $v_{\text{sc-so}}$=$|M(k)|^2/\epsilon^2\frac{D^{(0)}(q,\omega)}{1-|M(k)|^2D^{(0)}\Pi^0_g(q,\omega)/\epsilon}$ is the screened phonon exchange potential, $\Pi^0_{s}(q,\omega)\approx\frac{nq^2}{m^*\omega(\omega+i\delta_p)}$ is the polarizability for free carrier of GaAs, depending on $\delta_p$=$\hbar/\tau_p$. $v_c^{'}(q)=4\pi e^2/q^2$ is the 3D Coulomb potential of GaAs. From Eq. (\ref{eqn4}), we can obtain:

\begin{equation}
\frac{\epsilon_{\text{rpa}}}{\epsilon_{\text{eff}}}\approx1-\frac{\omega_{pl}^2(q)}{\omega(\omega+i\delta_e)}-\frac{2g^2\omega_{so}^2}
{\omega(\omega+i\delta_{so})-(1-2g^2)\omega_{so}^2}-\frac{\frac{\epsilon_{\infty}}{1+\epsilon_{\infty}}\omega_{p}^2}{\omega(\omega+i\delta_p)}
\label{eqn5}
\end{equation}
and the plasmon dispersion can also be obtained by using Eq. (\ref{eqn5}) with $\epsilon_{\text{rpa}}(q,\omega-i/\tau )=0$, and the carrier concentration dependent plasmon lifetime $\tau$ can be obtained from the imaginary part of plasmon energy.

\section{Carrier concentration dependent plasmon-phonon coupled modes}

\begin{figure}[htb]
\includegraphics[width=7cm]{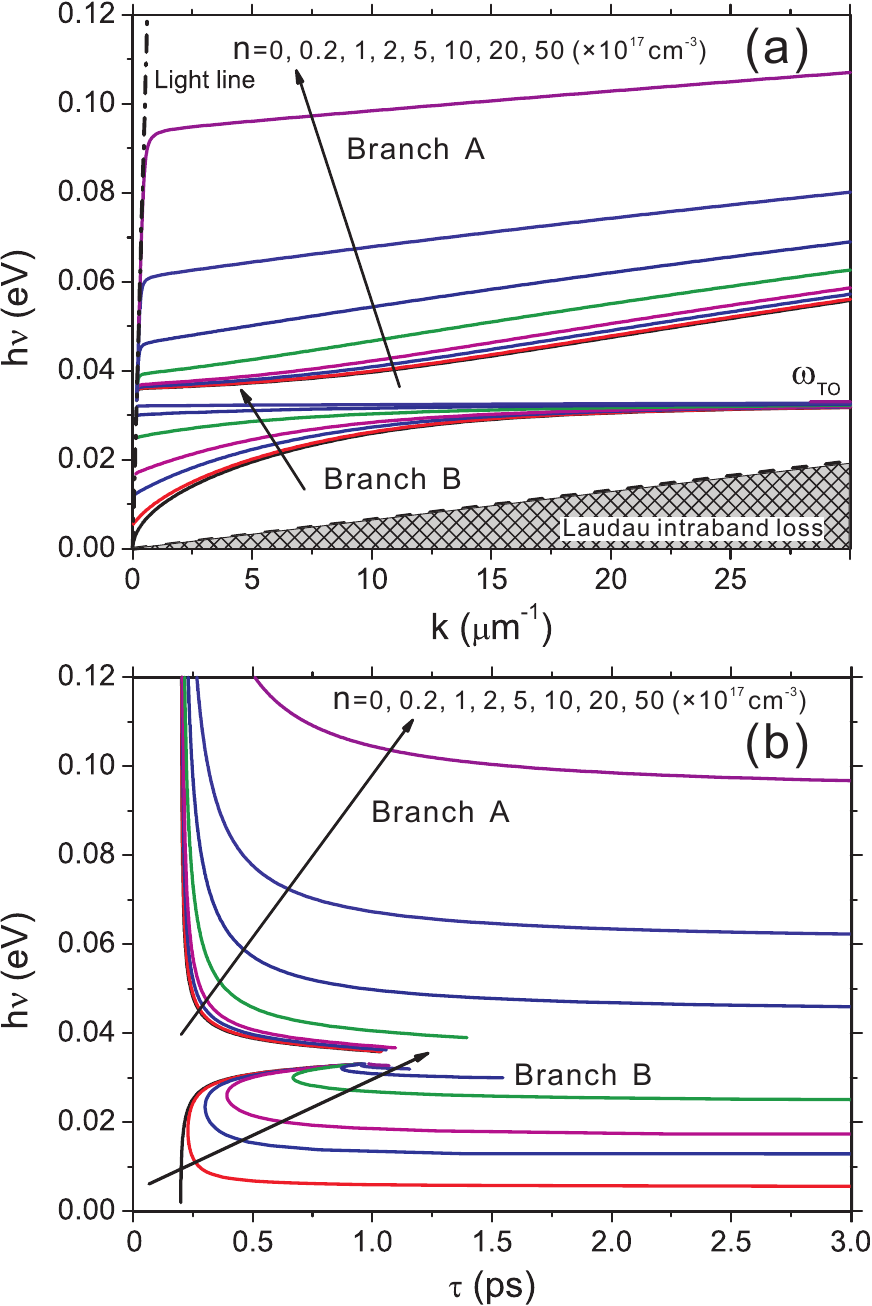}
\caption{\textbf{(a)} The dispersion of  plasmon-phonon coupled modes (PPCMs) for graphene lay on GaAs substrate with various carrier concentrations. The labels A and B of modes describe the coupled higher and lower energy branches, respectively. The black arrows mean the increasing of the carrier concentration in GaAs. The carrier concentrations are 0, $2\times10^{16}$, $1\times10^{17}$, $2\times10^{17}$, $5\times10^{17}$, $1\times10^{18}$, $2\times10^{18}$ and $5\times10^{18}$ $\text{cm}^{-3}$, respectively. The shadow triangle area indicates the Landau intraband loss. The parameters of graphene are assumed to be $E_f=0.2$ eV, and relaxation time $\tau_e=0.2$ ps (correspond to a DC mobility of 10000 $\text{cm}^2$/Vs) in the whole paper. \textbf{(b)} Plasmon lifetime of the two branches A and B with different carrier concentrations. The scattering of substrate is not included in the calculation. }
\label{fig2}
\end{figure}

First of all, the dispersion of PPCMs is calculated by using Eq. (\ref{eqn3}) without considering the damping of the subastrate ( $\Gamma$ and $\delta_p$ are set to 0) for simplicity. The carrier concentration dependent dispersion curves are shown in Fig. \ref{fig2}(a). It is worth noting that the intrinsic carrier concentration of GaAs is $n=2\times 10^{16}$ $\text{cm}^{-3}$. One can find that the dispersion of graphene plasmons splits into two branches due to plasmon-phonon coupling, and the two branches are labelled as A and B, respectively. First of all, for both branches A and B, the plasmon energy increases as the carrier concentration increases. Secondly,  there is a low energy cutoff frequency which is not existing in a normal polar material \cite{} for the branch B. Also the energy range of the branch B becomes narrower when the carrier concentration increases. Because the up limit $\omega_{\text{TO}}$ exists, the branch B only exists near $\omega_{\text{TO}}$, which leads to near-zero group velocity, this can be found applications in slow light propagation and optical storage. Thirdly, the branch A becomes nearly linear rather than quadratic (no coupling limit) when the carrier concentration is over $5\times 10^{17}$ $\text{cm}^{-3}$, which leads to group velocity dispersiveness GPs modes. It maybe important in communication applications.  Fig. \ref{fig2}(b) shows the lifetime of carrier concentration dependent PPCMs by using Eq. \ref{eqn5}. For simplicity, the free electron damping in GaAs is not considered (in coincide to the dispersion in Fig. \ref{fig2}(a)) and the surface-phonon damping $\tau_{so}$ is set as 1ps as reported in SiO$_2$ \cite{YLZ2013}. We can find that the plasmon lifetime becomes longer when the carrier concentration becomes larger.

\begin{figure}[htb]
\includegraphics[width=7cm]{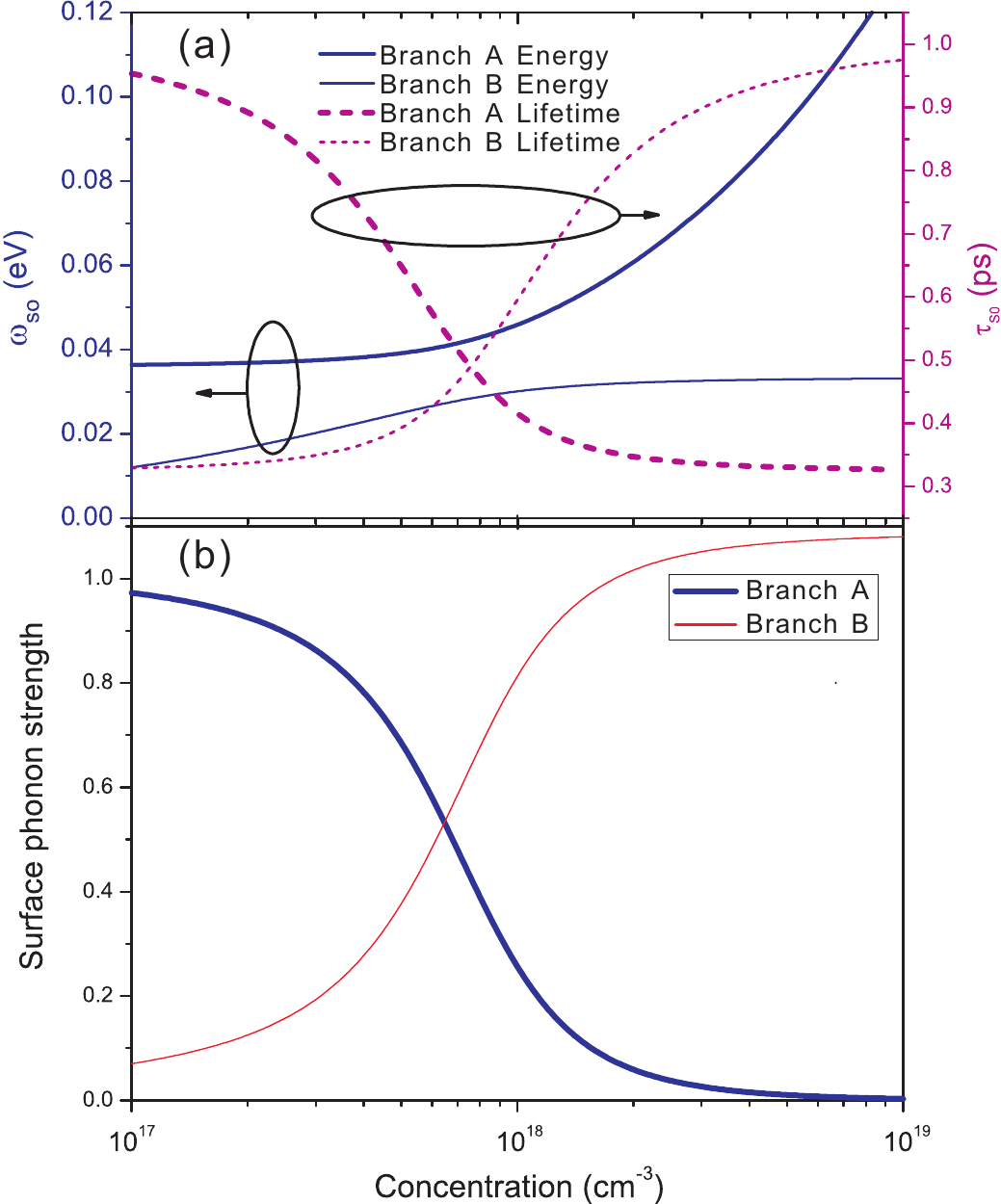}
\caption{\textbf{(a)} The energy and lifetime of suface phonon in doping GaAs, The thick(thin) line indicates the high(low) energy branch, and the dashed lines indicate their lifetime). \textbf{(b)} The surface phonon strength of the branches according to Fig. \ref{fig3}(a).}
\label{fig3}
\end{figure}

To understand the dispersion and lifetime of PPCMs, the coupled surface phonon of the substrate is introduced and described, which is calculated by $(\epsilon(\tilde{\omega}_{so}-i/\tilde{\tau}_{so})+1)/2=0$ with Eq. (\ref{eqn2}). These results are shown in Fig. \ref{fig3}(a). One can find there are two branches for $\tilde{\omega}_{so}$. The high and low energies $\tilde{\omega}_{so}$ are the approximate cutoff frequencies for branches A and B shown in Fig. \ref{fig2}(a), respectively. In further, one can find that the cutoff frequency of branch A monotonously increases from the origin value $\omega_{so}$ and does not have up limit in our calculated carrier concentration range, while the cutoff frequency of branch B monotonously increases from 0 (no cutoff frequency) to $\omega_{\text{TO}}$. Moreover, the energy of branch A surpasses the high energy $\tilde{\omega}_{so}$, and the energy of branch B locates between the low energy $\tilde{\omega}_{so}$ and $\omega_{\text{TO}}$. Thus, one can know that the behavior of  carrier concentration dependent PPCMs shown in Fig. \ref{fig2}(a) results from the controlling of the surface phonon of substrate.

From the surface phonon strength definition $S_m=|<m|\phi_k|0>|^2$ , where $|m>$ is the one-phonon excited state in the $\textit{m}$th level, $\phi_k=b_{-k}+b_{k}^{\dagger}$ and $b_{k}$ is surface phonon creation operator. Surface phonon strength is given by \cite{}

\begin{equation}
S_m=\frac{\sqrt{\epsilon_{\infty}/\epsilon_0}(\epsilon_{\infty}/\epsilon_0-1)x_m^3}{(\epsilon_{\infty}/\epsilon_0)x_m^4+y^2(1-x_m^2)^2}
\label{eqn6}
\end{equation}
where $x=\tilde{\omega}_{so}/\omega_{\text{TO}}$ is the normalized coupled surface phonon energy, and $y=\omega_{p}/\omega_{\text{TO}}$ is normalized bulk plasmon of GaAs. The value $S_m$ describes the weight of the two surface phonons, which contributes to the property of total surface phonons. The strength of surface phonons is calculated and shown in the Fig. \ref{fig3}(b), we know that the phonon strength of branch A (B) decreases (increases) dramatically as the carrier concentration increases, similar to the behavior of phonon lifetime shown in Fig. \ref{fig3}(a).

\begin{figure}[htb]
\includegraphics[width=7cm]{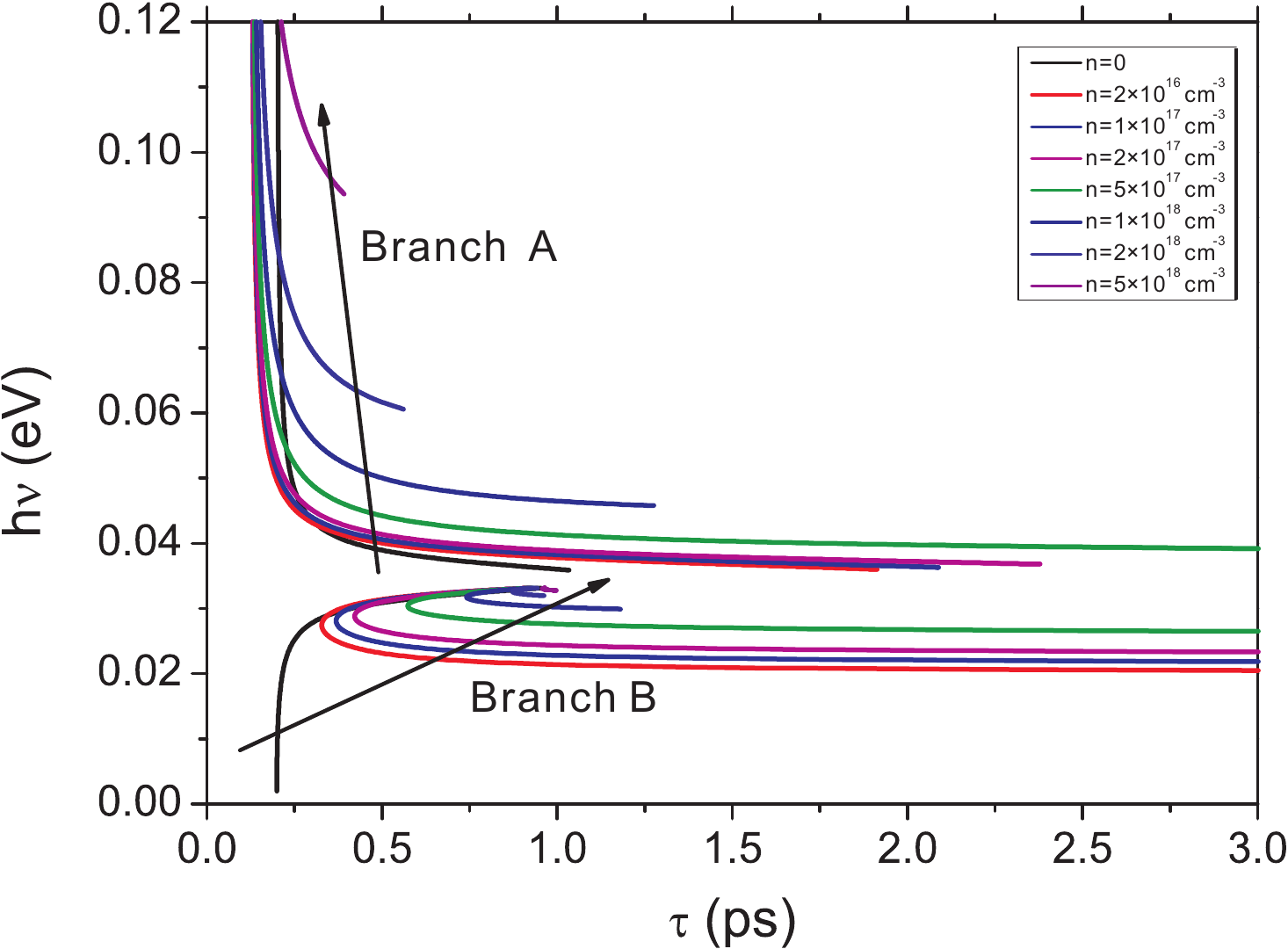}
\caption{The plasmon lifetime of the coupled plasmon-phonon modes with the electron scattering of the substrate GaAs. It is notable that the curve $n=0$ is the same as it in Fig. \ref{fig2}(b).}
\label{fig4}
\end{figure}

To further understand the lifetime of PPCMs, in Fig. \ref{fig2}(a),  the lifetime does not include the contribution from the scattering of free electrons in GaAs. More specifically, the lifetime of surface phonon is set to a constant value 1ps in our pervious calculations. However, from Fig. \ref{fig3}(a) we know that the lifetime of coupled surface phonon will decrease (increase) for branch A (B) with electron scattering. So the electron scattering effect must be considered in further calculations if one wants to compare the theoretical results with actual experiments. By substituting the $\delta_p$ into Eq. (\ref{eqn5}), the total lifetime of GPs is calculated and shown in Fig. \ref{fig4}. The upper limit for the branch A of PPCMs decreases apparently. But the upper limit for the branch B of PPCMs is long enough for proper carrier concentration.

\begin{figure}[htb]
\includegraphics[width=7cm]{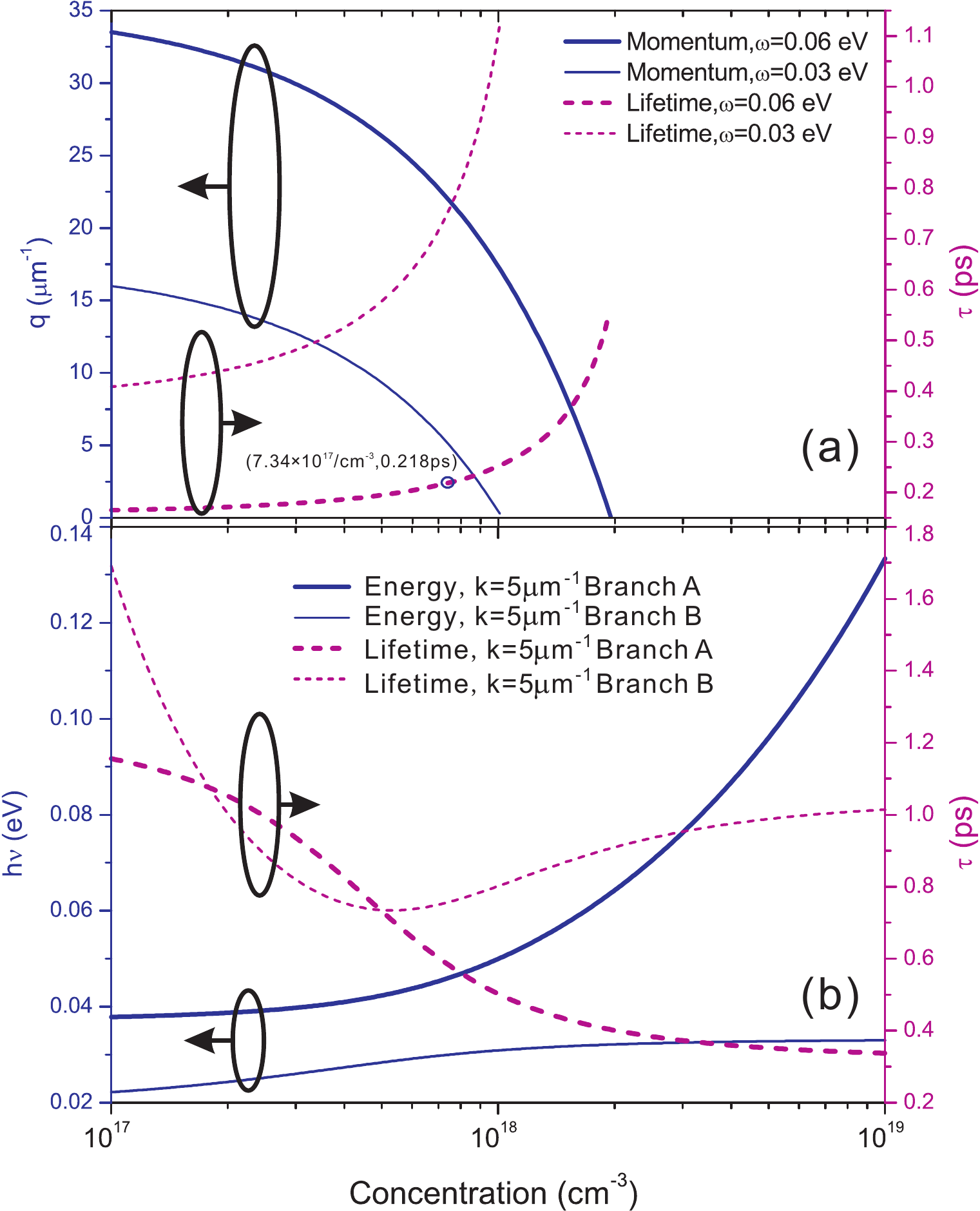}
\caption{\textbf{(a)} Carrier concentration dependent plasmon momentum and lifetime for given photon energies. The thick line ($\hbar\omega=0.06$ eV) is chosen from branch A, and the thin line ($\hbar\omega=0.03$ eV) from branch B. The dashed lines show the corresponding lifetime of the branches. \textbf{(b)} Concentration dependent energy for a given momentum 5 $\mu m^{-1}$, the thick line indicates the branch A and the thin line indicates the branch B.}
\label{fig5}
\end{figure}

For actual experiments, we usually have a given light source and a given momentum ($\sim1/a$), $a$ is the structure size for a tip or a grating, because the efficient excitation of GPs always requires to fulfill momentum conservation conditions. So finite discrete wavelengths and momentums are taken to realize GPs excitation and propagation. First, we analyze the dispersion of PPCMs with a given wavelength. Without loss of generality, two energies from branch A ($\hbar\omega=0.06$ eV) and branch B ($\hbar\omega=0.03$ eV) are chosen. The momentum and lifetime of the carrier concentration dependent PPCMs are displayed in Fig. \ref{fig5}(a). One can find that the momenta of branches A and B (the solid lines) decrease to zero as the concentration increases. The result indicates that for a given monochromatic wave, we can always find a proper concentration to match the momentum out of the light cone. This effect opens a door for that all kinds of confined modes or evanescent waves can be coupled to PPCMs. The corresponding lifetime of these branches is indicated by the dashed lines. And the lifetime increases dramatically as the momentum decreasing. This is due to the effect of phonon coupling and low mode confinement. The critical point showed in Fig. \ref{fig5}(a) is the lifetime of zero concentration position. Because of the absence of GaAs electron scattering damping path, it is longer than low concentration condition with electronic damping. When the concentration is larger than $8.12\times10^{17}$ $\text{cm}^{-3}$, the lifetime of PPCMs can be longer than zero concentration lifetime for $\hbar\omega=0.06$ eV ($\tau=0.203$ ps). For the energy $\hbar\omega=0.03$ eV the lifetime is always larger than zero concentration lifetime ($\tau=0.215$ ps) in our calculation carrier concentration range. On the other hand,  for a given momentum $q=$ 5 $ \mu m^{-1}$, the PPCMs energy and lifetime are shown in Fig. \ref{fig5}(b). From the figure, we know that the branches A and B show similar behavior in the case with $\hbar\omega_{so} (q\to \infty)$ (Fig. \ref{fig3} (a)). If a wide spectrum pulse is given, one can tune the output frequency of PPCMs by changing the carrier concentration in a fixed experimental scheme.

\begin{figure}[htb]
\includegraphics[width=7cm]{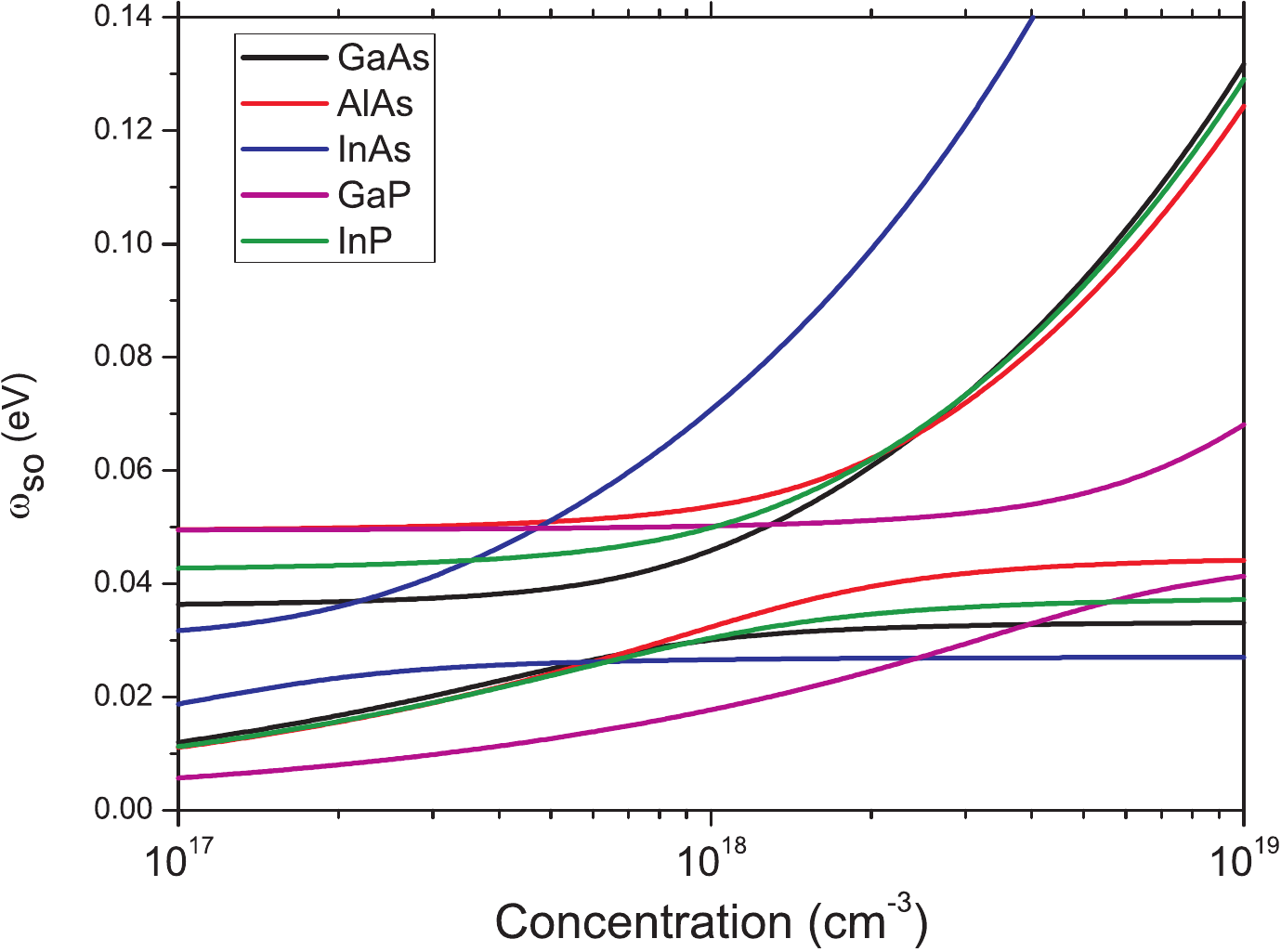}
\caption{Carrier concentration dependent surface phonon energy in different semiconductors, GaAs, AlAs, InAs, GaP and InP.}
\label{fig6}
\end{figure}

Finally, coupled surface phonon energy change with carrier concentration for several semiconductor materials is inverstigated. The change range is determined mainly by the effective carrier mass of semiconductor. For InAs, surface phonon energy can reach $0.2$ eV for concentration near $10^{19}$ $\text{cm}^{-3}$, which means that all the energy below intrinsic graphene phonon can lead to PPCMs.

\section{Conclusions}
In summary, in this paper we investigated analytically the effect of substrate carrier concentration on the properties of PPCMs. By using the semiconducor substratrate GaAs,  we found the coupled modes can be effectively controlled by changing the carrier density of the substrate. Specifically, the dispersion and lifetime of PPCMs can be controlled by the carrier density of GaAs. In further, the effect can be understood by the surface phonon of substrate change. The controllable long-live and easily excited PPCMs can find applications using graphene plasmons.

\section{acknowledgments}

This work was financially supported by the National Basic Research Program of China (2010CB934101, 2013CB328702), the National Natural Science Founda- tion of China (11374006, 11304162), the 111 Project (B07013).

%\begin{thebibliography}{99}
%\bibitem{web} \url{http://www.ecse.rpi.edu/~schubert/Course-reference-materials/Materials-Semiconductors-Si-Ge-GaAs-&-GaN.pdf}.
%\end{thebibliography}

%\bibliographystyle{aipnum4-1}
%\bibliography{ref}
%

\end{document}